\documentclass[aps,prc,twocolumn,10pt, superscriptaddress, showpacs, floatfix]{revtex4-1}

\usepackage{amssymb,epsfig}
\usepackage{bbold}

\newcommand{\be}{\begin{equation}}
\newcommand{\ee}{\end{equation}}
\newcommand{\bea}{\begin{eqnarray}}
\newcommand{\eea}{\end{eqnarray}}

\begin{document}

\title{Relativistic approach to the nuclear breathing mode}

\author{Elena Litvinova} 
\affiliation{Department of Physics, Western Michigan University, Kalamazoo, MI 49008, USA}
\affiliation{National Superconducting Cyclotron Laboratory, Michigan State University, East Lansing, MI 48824, USA}
\date{\today}

\begin{abstract}
Microscopic theory of the nuclear response based on the relativistic meson-nucleon Lagrangian is applied to the description of the isoscalar giant monopole resonance (ISGMR) in a variety of nuclear systems. It is shown that the parameter-free inclusion of beyond-mean-field correlations of the quasiparticle-vibration coupling (qPVC) type in the leading approximation allows for a simultaneous realistic description of the ISGMR in nuclei of lead, tin, zirconium, and nickel mass regions, which is difficult on the mean-field level. The calculations 
employ the finite-range effective meson-nucleon interaction, which, in combination with the qPVC, has consistently demonstrated the ability to reliably describe many other nuclear structure phenomena. Systematic calculations of the isoscalar monopole response for nickel isotopes help reveal the central role of the coupling between the ISGMR and the low-energy quadrupole states in the placement of the ISGMR centroids.
\end{abstract}

\maketitle

%

The nuclear compression modes of excitation, first of all, the isoscalar giant monopole resonance (ISGMR), remain in the focus of 
experimental and theoretical research since decades. Probing the nuclear response to small density fluctuations allows extracting
the nuclear compression modulus, also known as nuclear incompressibility $K_{\infty}$. Because of the direct link between the centroid of the ISGMR and $K_{\infty}$, 
the ISGMR provides fundamental information about  the nuclear equation of state 
as the nuclear incompressibility governs the density fluctuations around the saturation point of the nuclear matter.

The major experimental tool to study the ISGMR is an inelastic scattering of alpha particles or deuterons to small forward angles. The techniques developed and implemented in RCNP at Osaka University, Texas A$\&$M University Cyclotron Institute and KVI in Groningen have allowed collecting comprehensive information on the ISGMR gross structure in a variety of nuclear systems, see review \cite{Garg2018}. Recent experiments at iThemba LABS were focused on revealing fine details of the ISGMR, which are very instructive for understanding microscopic mechanisms of its formation \cite{Olorunfunmi2022,Bahini2022}. The most powerful theoretical approach is the nuclear response theory. Its simplest version valid for open-shell nuclei, the quasiparticle random phase approximation (QRPA), was extensively applied to the ISGMR in the context of its simultaneous description in a variety of medium-heavy nuclei within a single approach. Most of the recent studies employed effective interactions of the density functional theory (DFT) \cite{Piekarewicz2007,Khan2010,Kvasil2016,Sun2021}. It turned out, in particular, that the  DFT parametrizations with commonly accepted $K_{\infty}$ values describing well the ISGMR in $^{208}$Pb notably overestimate the ISGMR centroid in tin isotopes \cite{Garg2018}. Some attempts to solve this problem by including correlations beyond QRPA were undertaken by non-relativistic approaches based on the Skyrme interaction, however, they either employed a non-self-consistent calculation scheme \cite{Tselyaev2009} or 
implied fine-tuning of the underlying DFT \cite{Li2022}.


In Ref. \cite{LitvinovaTrentoGMR2022} I presented a conjecture that this problem may have a  beyond-QRPA solution within a self-consistent relativistic approach, which does not require a specially designed parametrization, but instead is based on the known parameter set and already has a record of a reasonable description of other nuclear structure phenomena. 
In the present article, this idea is elaborated in a framework rooted in the effective meson-nucleon Lagrangian and extended beyond the relativistic QRPA (RQRPA) solutions. 
The general formalism is given in Ref. \cite{LitvinovaZhang2022}, where the equation of motion (EOM) for the response of a superfluid fermionic system derived from the bare nucleonic Hamiltonian is shown to have an interaction kernel, which is naturally split into the static and dynamical parts. The former governs the short-range correlations, while the latter is responsible for the long-range ones. The dynamical kernel heading beyond the QRPA-type EOM is generally coupled to a hierarchy of higher-rank correlation functions, however, this hierarchy can be truncated at different levels. In the leading approximation, the dynamical kernel can be reduced to the amplitude of the quasiparticle-vibration coupling (qPVC), where the Bogoliubov quasiparticles are minimally coupled to the superfluid phonons. 

While ab-initio implementations of the latter approach remain a task for the future, a self-consistent scheme has been developed for calculations with effective interactions derived from the energy density functionals. In such a scheme, (i) the static kernel as a whole is replaced by the effective interaction, (ii) the bare interaction in the dynamical kernel is replaced by the effective interaction with (iii) simultaneous subtraction of the static limit of the dynamical kernel \cite{Tselyaev2013}, and (iv) the phonons entering the dynamical kernel can be calculated at the (R)QRPA level.
\begin{figure*}
\begin{center}
\includegraphics[scale=0.57]{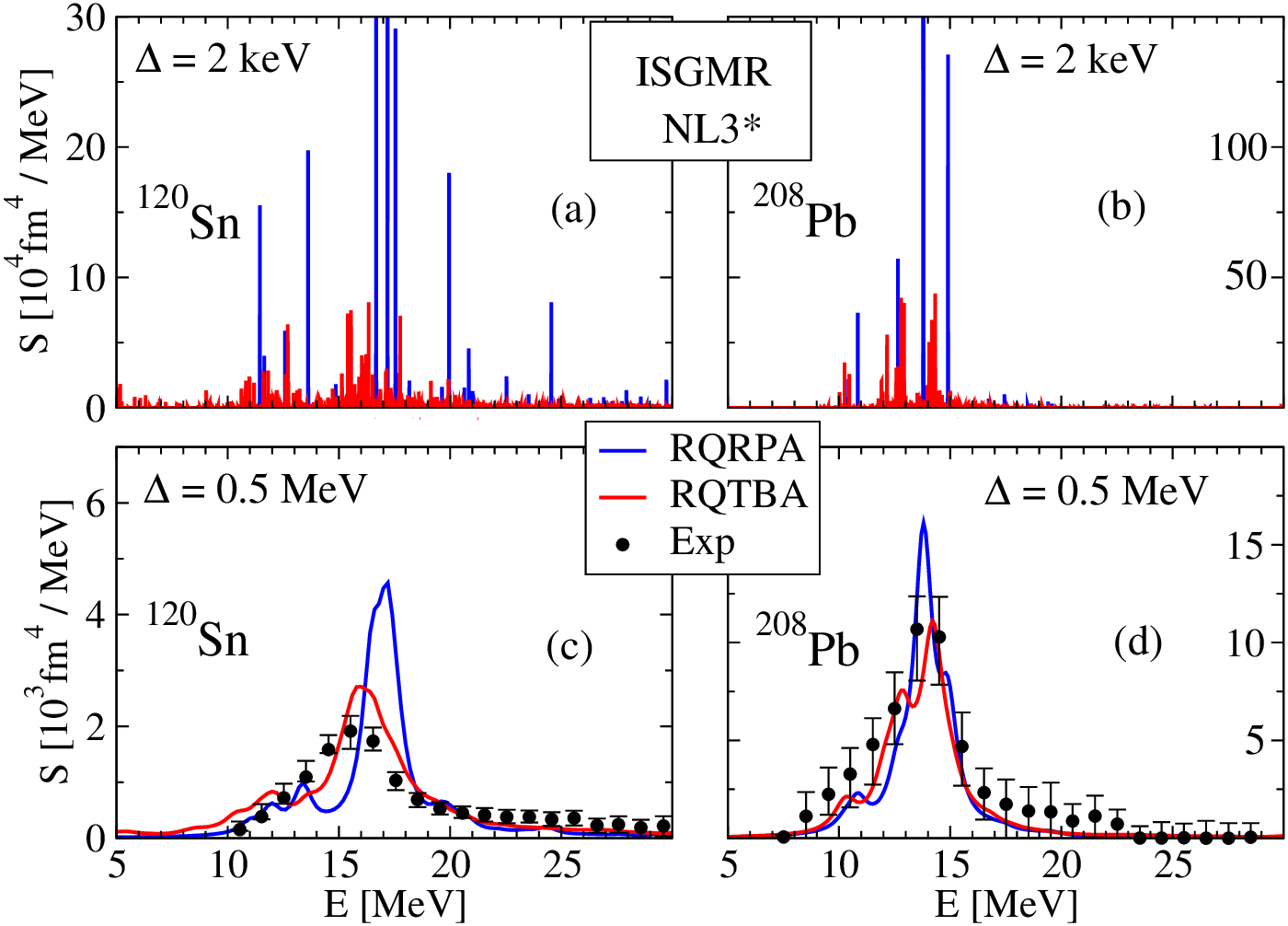}
\end{center}
\caption{ISGMR in $^{120}$Sn and $^{208}$Pb: RQRPA and RQTBA strength distributions compared to experimental data \cite{Li2007} ($^{120}$Sn) and \cite{Garg2018} ($^{208}$Pb). }
\label{ISGMR}%
\end{figure*}

In the relativistic framework, this approach was pioneered in Refs. \cite{LRT2007,LRT2008} as the relativistic quasiparticle time blocking approximation (RQTBA) including the qPVC in a parameter-free way. 
Later on, it was re-derived in a more fundamental EOM formalism based on the bare fermionic Hamiltonian for the two-point response function, both normal \cite{LitvinovaSchuck2019} and superfluid \cite{LitvinovaZhang2022}.
This theory does not require time blocking and allows for consistent extensions beyond the leading approximation, however, the name RQTBA is retained as the DFT-based implementation remains valid.
Both the original and extended \cite{LitvinovaSchuck2019} versions of the RQTBA have shown a good performance in the description of a variety of nuclear excited states, in both neutral \cite{LRT2008,EndresLitvinovaSavranEtAl2010,LanzaVitturiLitvinovaEtAl2014} and charge-exchange \cite{RobinLitvinova2016,RobinLitvinova2018,Robin2019,Scott2017} channels.  In particular, the RQTBA demonstrated remarkable improvements with respect to the RQRPA. Systematically, the configurations with two quasiparticles coupled to a phonon ($2q\otimes phonon$) generated by the qPVC produce a reasonable degree of fragmentation of the pure $2q$-states already in the leading approximation. Furthermore, the description of the low-energy (soft) modes was refined considerably. That is especially important for astrophysical applications, such as the r-process nucleosynthesis and the supernovae evolution \cite{LRW2020,LR2021}.   

As in the applications listed above, in this work the calculations were performed with the non-linear sigma-meson parametrization of third generation. The eight-parameter set NL3$^{\ast}$ \cite{Lalazissis2009} with the compression modulus $K_{\infty}$ = 258 MeV is an improved version of the NL3 one with the larger $K_{\infty}$ = 272 MeV \cite{Lalazissis1997}. Both the NL3 and NL3$^{\ast}$ imply the finite-range meson-exchange forces compatible with the PREX data on the neuron skin thickness in $^{208}$Pb \cite{Adhikari2021} (note that Ref. \cite{Lalazissis2009} gives an erroneously small value for the neutron skin of $^{208}$Pb) and showed similar results in most applications on the mean-field and RQRPA levels as well as beyond them. For calculations of the ISGMR, which is directly linked to the nuclear compressibility, the NL3$^{\ast}$ is a better choice because of its more realistic value for the $K_{\infty}$. The more recent point-coupling relativistic parametrizations require smaller $K_{\infty}$ = 230 MeV just because of their zero-range character, however, they provide a similar quality of description of many nuclear structure phenomena including the nuclear monopole response \cite{Niksic2008}, so that the use of the NL3$^{\ast}$ forces in this work is well justified. 

\begin{figure*}
\begin{center}
\includegraphics[scale=0.57]{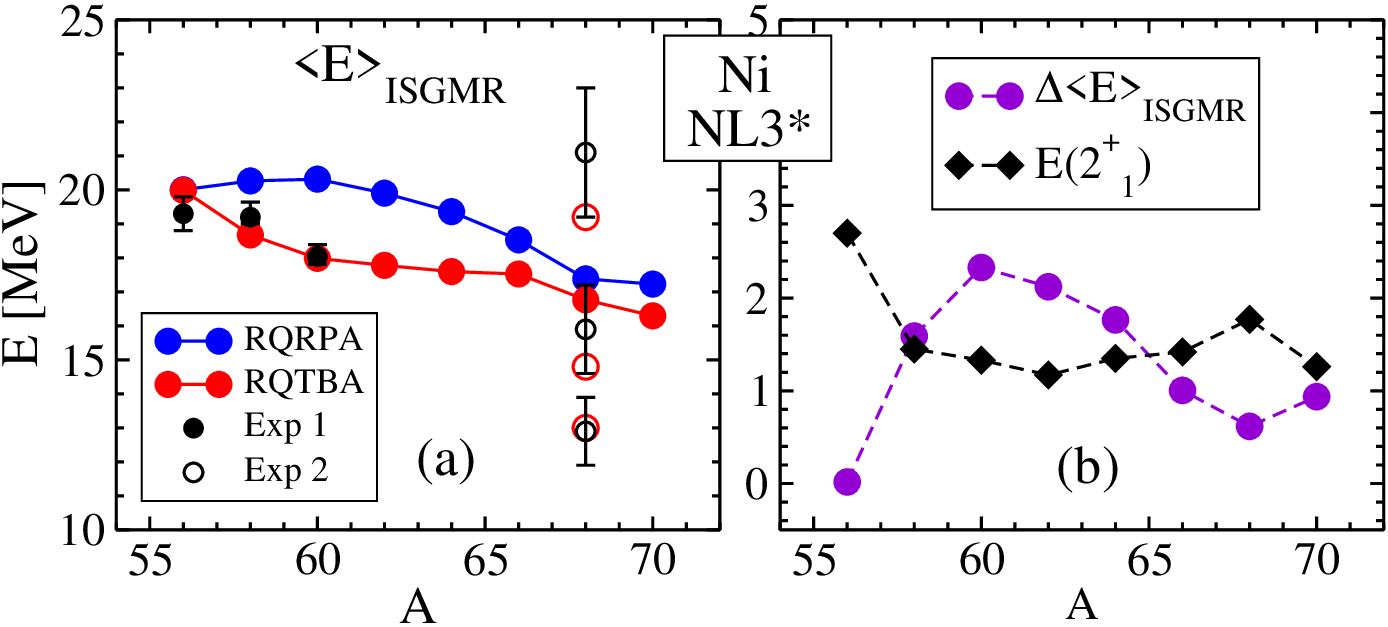}
\end{center}
\caption{Panel (a): ISGMR centroids in nickel isotopes $^{56-70}$Ni compared to data of Refs. \cite{Monrozeau2008} ($^{56}$Ni), \cite{Lui2006} ($^{58,60}$Ni), and \cite{Vandebrouck2014} ($^{68}$Ni).  Empty symbols stand for three separate peaks above 10 MeV determined in \cite{Vandebrouck2014}. Panel (b): the downward shifts $\Delta\langle E \rangle_{\text{ISGMR}}$ of the ISGMR centroids in the RQTBA with respect to the RQRPA ones (circles) in comparison with the E(2$^{+}_1$) values (diamonds).}
\label{ISGMRc}%
\end{figure*}

Fig. \ref{ISGMR} shows the ISGMR in the $^{120}$Sn and $^{208}$Pb nuclei calculated within the self-consistent RQRPA based on the development of Ref. \cite{Paar2003} and the RQTBA including the qPVC in a parameter-free way \cite{LRT2008}. Here the advantage that the most important low-lying phonons are well reproduced in the RQRPA \cite{LR.2006,AL.2015} is employed. The phonon model space included the RQRPA modes of natural parity up to spin $J = 6$ and energy 20-30 MeV, and was slightly truncated on the coupling strength using the same criteria as in the series of earlier calculations, for instance, in Ref. \cite{EL2016}. The subtraction procedure applied following Ref. \cite{Tselyaev2013} ensures converging results within the so truncated configuration space. 
The superfluid pairing correlations were accurately taken into account within the Bogoliubov's theory for the ground state and, respectively, in the qPVC amplitude of the RQTBA for the excited states. The use of the monopole pairing forces has the advantage of equivalence between the Bardeen-Cooper-Schrieffer and the Bogoliubov approaches to the superfluid pairing. The flexibility of the pairing interaction strength was used to slightly adjust the pairing gaps to their empirical values. 

The calculations were performed for the smaller ($\Delta =$ 2 keV)  and larger ($\Delta =$ 0.5 MeV) values of the smearing parameter commonly used to smooth the strength distributions $S(E)$ as defined, e.g., in Refs. \cite{LRT2008,LitvinovaZhang2022}. As this parameter equals to the half width of the resulting Lorentzian peaks \cite{LitvinovaZhang2022}, the latter value of the smearing parameter corresponds to the energy resolution of the experimental data $\approx$1 MeV, compared to the calculations in panels (c, d). One can see that the RQRPA provides a good description of the ISGMR centroid in $^{208}$Pb, while it is positioned too high in $^{120}$Sn. 
The inclusion of the qPVC in the RQTBA famously introduces a considerable fragmentation of the RQRPA modes, as expected. It can be seen that, in addition, the ISGMR centroid in $^{120}$Sn moves to lower energy, while the one in $^{208}$Pb remains practically intact. In both cases, a visible improvement of the ISGMR description is achieved in the RQTBA.
Note again that the qPVC here was included in the leading approximation, and the theory is shown to be systematically improvable while, at the same time, converging \cite{LitvinovaSchuck2019,LitvinovaZhang2022}.
Panels (a,b) illustrate the fine structure of the ISGMR, its fragmentation and overall descent due to the qPVC in both nuclei, providing the details not visible with large smearing.
  
What makes the qPVC effects in  $^{120}$Sn and $^{208}$Pb different? While $^{208}$Pb is a canonical doubly-magic, or closed-shell, nucleus, $^{120}$Sn is a typical open-shell one. A number of earlier studies of other structure phenomena in medium-heavy nuclei, for instance, the dipole excitation modes  \cite{LRT2007,LRT2008} or the single-quasiparticle states \cite{LR.2006,AL.2015}, points out that the stronger qPVC in open-shell nuclei has a systemic character. Ref. \cite{Wibowo2022} studied the qPVC mechanism within toy models with variable parameters and indicated, in particular, that moving the phonons to lower energies enhances the qPVC effects. 
Another essential factor is the qPVC coupling strength, which increases with the phonon collectivity. From numerous realistic calculations it follows that the most important role is played by the low-spin, first of all, the quadrupole and octupole phonons. Finally, knowing that the ISGMR is strongly coupled to quadrupole modes \cite{Garg2018} narrows the argumentation down to the energies of the lowest collective quadrupole states E(2$^{+}_1$). 
At this point, one can note that E(2$^{+}_1$;$^{208}$Pb) = 4.09 MeV $\gg$ E(2$^{+}_1$;$^{120}$Sn) = 1.17 MeV, which supports the conjecture about the stronger qPVC in $^{120}$Sn. However, these two nuclei are different in many aspects, so that this argument alone may be not conclusive. A better option could be to examine a set of nuclei with similar particle content, for which the ISGMR data are available and which could, thus, be the tin isotopic chain $^{112-124}$Sn. Remarkably, the ISGMR in these nuclei looks very similar  \cite{Li2007,Tselyaev2009} while, accordingly, the E(2$^{+}_1$) values  and the corresponding transition probabilities are very close \cite{nndc}, however, the same trend makes it difficult to discriminate the influence of the qPVC to the 2$^{+}_1$ state on the ISGMR among these isotopes.

Obviously, it is more instructive to consider  isotopes or isotones with significantly varying E(2$^{+}_1$). In this context, a good case can be the chain of stable nickel isotopes, for which some ISGMR data are available.
Fig. \ref{ISGMRc} illustrates the ISGMR centroids computed in the RQRPA and RQTBA in comparison to data (a) and shifts of these centroids due to the qPVC in comparison with the E(2$^{+}_1$) values (b) in the even-even nickel isotopes $^{56-70}$Ni. 
The centroid was defined as the ratio of the first and zeroth moments $\langle E \rangle_{\text{ISGMR}} = m_1/m_0$ of the obtained ISGMR strength functions $S(E)$ in the energy interval $0\leq E \leq 30$ MeV. 

The RQRPA centroid energy (blue symbols in the panel (a) of Fig. \ref{ISGMRc}) slightly increases when moving from the doubly closed-shell $^{56}$Ni to the open-shell $^{58}$Ni and then shows a smooth descent with the increase of the neutron number. A kink seen in $^{68}$Ni is associated with the strong subshell closure $N = 40$. This trend occurs mostly due to the pairing correlations, which are absent in $^{56}$Ni, maximized in mid-shell isotopes, and then suppressed in $^{68}$Ni. The red symbols in the same panel are assigned to the RQTBA results showing how the ISGMR centroids are modified by the qPVC. Their values are going down revealing the biggest shifts in mid-shell nuclei and again a little kink at $^{68}$Ni. A direct comparison to the experiment is possible for $^{56,60,62}$Ni, and the agreement is quite reasonable for the RQTBA centroids. 
In $^{68}$Ni, the data available from Ref. \cite{Vandebrouck2014} are given for the positions of three peaks identified above 10 MeV, therefore, the energies of the corresponding peaks in RQTBA are also plotted for comparison, and found to be within the experimental error bars. 

Panel (b) of Fig. \ref{ISGMRc} displays the values of the downward shifts $\Delta\langle E \rangle_{\text{ISGMR}}$ of the ISGMR centroids in RQTBA with respect to the RQRPA ones in comparison with the E(2$^{+}_1$) values in the same chain of nickel isotopes. Here the experimental E(2$^{+}_1$) values are plotted, while the theoretical lowest quadrupole phonon energies are rather close to that values as they are used for fine-tuning the pairing gaps. One can see that the energies E(2$^{+}_1$) are in a clear countertrend with $\Delta\langle E \rangle_{\text{ISGMR}}$: the decrease of the  E(2$^{+}_1$) values corresponds to the increase of the ISGMR centroid shifts $\Delta\langle E \rangle_{\text{ISGMR}}$ and vice versa, thereby confirming the initial conjecture. Note here that the decrease in  E(2$^{+}_1$) is typically associated with the increase of the collectivity of the quadrupole transitions reflected by the transition probabilities, and this correlation is especially clear in nickel isotopes \cite{Sorlin2002}. Thus, the ISGMR centroid shift is the direct consequence of the quadrupole collectivity.  

In certain nuclei, this type of collectivity tends to induce a static deformation, and this is the case for $^{60,62}$Ni. RQRPA calculations, which allow for an axial deformation, using the approach of Ref. \cite{Bjelcic2020} were performed to test the sensitivity of the ISGMR to deformations, and the ISGMR centroids remained robust within a few hundred keV.
This result agrees with that of Ref. \cite{Kvasil2016} obtained with the Skyrme interaction, which indicates a weak sensitivity of the ISGMR to deformations.
Thus, the RQTBA calculations for these nuclei can be done in a spherical basis, while the static quadrupole deformation is modeled by the coupling to the lowest quadrupole phonon. 

%
\begin{figure}
\begin{center}
\vspace{0.5cm}
\includegraphics[scale=0.37]{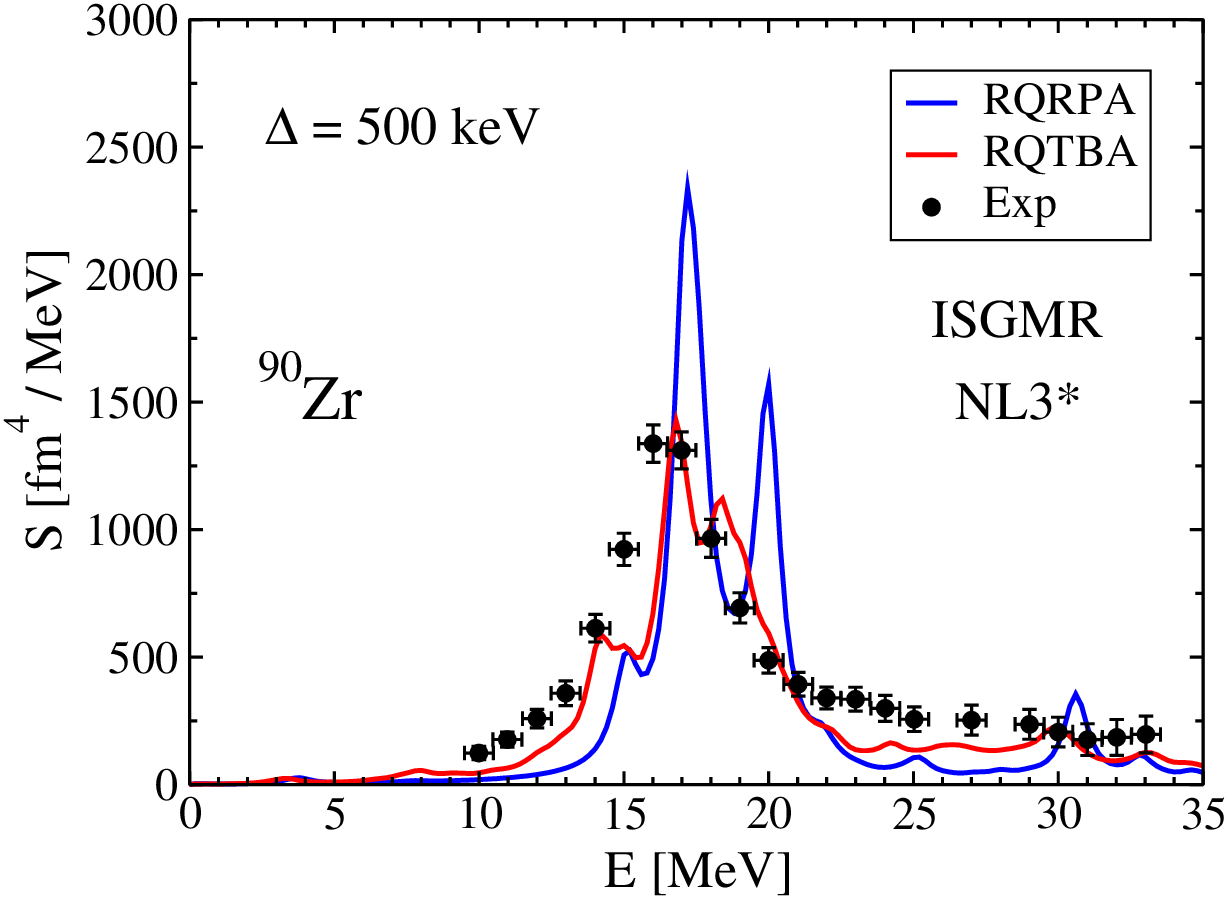}
\end{center}
\caption{ISGMR in $^{90}$Zr: RQRPA and RQTBA strength distributions compared to experimental data \cite{Gupta2016}.}
\label{90zr}%
\end{figure}

Among spherical nuclei, $^{90}$Zr constitutes an interesting case, which is somewhat borderline between $^{208}$Pb and mid-shell tin isotopes discussed above.
Specifically, this nucleus is known to present no fluffiness
puzzle, i.e., properties consistent with those of $^{208}$Pb, with respect to the
relationship between the GMR energy and the compression modulus. At
the same time, the experimental energy of its first $2^+$ state is only 2.19 MeV, which is not as high as
in $^{208}$Pb. This energy is higher than in its neighboring isotopes, which is the consequence of the neutron $N = 50$ shell closure, while the proton number $Z = 40$ is associated with a relatively strong closed subshell. The ISGMR obtained for $^{90}$Zr within the same calculation scheme is displayed in Fig. \ref{90zr}, in comparison with the data of Ref. \cite{Gupta2016}. One can see that the qPVC taken into account in the RQTBA in the leading approximation, with the phonon characteristics obtained from the RQRPA, is capable to improve both the width and the centroid of the ISGMR, as compared to the RQRPA, also in this nucleus. A little more modest shift of the ISGMR centroid than in $^{120}$Sn is obtained for $^{90}$Zr, which is consistent with the trend of the E(2$^{+}_1$) values. 

As it is briefly noted above, although the leading-approximation qPVC  improves considerably the description of the ISGMR as well as of a wide range of other nuclear structure phenomena, compared to the (R)QRPA and mean-field models, it does not provide a perfect description. Being an essentially non-perturbative emergent phenomenon, the qPVC further induces higher-complexity collective correlations, including the ground-state correlations, some of which were addressed in Refs. \cite{LitvinovaSchuck2019,Robin2019}. While the leading-approximation qPVC controls mostly the gross nuclear spectral properties, a consistent inclusion of the higher-complexity correlations, along with the novel highly-accurate DFTs and bare nucleon-nucleon interactions, has the potential of resolving their finer details.

In summary, 
the idea of a description of the nuclear monopole response, or breathing mode, in medium-heavy nuclei was elaborated within a relativistic many-body approach based on the universal eight-parameter set and taking into account the qPVC in the leading approximation in a parameter-free way. It was shown that, although the placement of the ISGMR centroid generally depends on the nuclear compression modulus, it is further fine-tuned naturally by the coupling of the ISGMR to the low-energy phonons, mostly those of quadrupole character. This coupling causes spreading and an overall shift of the ISGMR centroid down with respect to its value obtained in the RQRPA, which can amount up to 1-2 MeV and which is more pronounced in softer mid-shell nuclear species, in agreement with data. It is noted that the same approach has led to a systematic improvement of the description of other nuclear structure properties as compared to the commonly adopted DFT.
The obtained result, thus, further highlights the insufficiency of the DFT description of complex nuclear phenomena and the necessity of consistently including complex correlations, in particular, in the determination of the nuclear equation of state.

This work was supported by the US-NSF CAREER Grant PHY-1654379 and US-NSF Grant PHY-2209376.


\end{document}